\documentclass{aa}

\usepackage{epsfig,latexsym,longtable,lscape,supertabular}
\begin{document}
\pagenumbering{arabic}

\title{The Cartwheel galaxy with XMM-Newton}
\subtitle{}

\author{Erika Crivellari, Anna Wolter
\and Ginevra Trinchieri}

   \offprints{A. Wolter: anna.wolter@brera.inaf.it}

\institute{INAF-Osservatorio Astronomico di Brera, via Brera 28, 20121 Milano, Italy}
%\and <name of the second institute> ...}

\date{Received ../ Accepted }

\abstract {} 
% Aims
{The extreme environment provided by the Cartwheel ring is analyzed to study
its X-ray and optical-UV properties. We compare the Cartwheel
with the other members of its group and study the system as a whole 
in the X-ray band.}
% Methods 
{We analyze the data of the Cartwheel galaxy obtained
with \emph{XMM-Newton} in two different periods (December 2004 and May
2005). We focus on the X-ray properties of the system and  use the
OM data to obtain additional information in the optical and UV
bands. Each dataset is analyzed separately to study source variability 
and summed together to study fainter and extended sources. }
% Results
{We detect a total of 8 sources associated with the Cartwheel galaxy
and three in its vicinity, including G1 and G2, all at L$_X \geq 10^{39}$ erg
s$^{-1}$, that is the Ultra Luminous X-ray (ULX) source range.
The brightest ULX source has been already discussed elsewhere.
The spectra of the next three brightest ULX are well 
fitted by a power-law model with a mean photon index
of $\sim2$.
We compare the \emph{XMM-Newton} and \emph{Chandra} datasets to study 
the long-term variability of the sources.  
At least three sources vary in the 5 months
between the two \emph{XMM-Newton} observations and 
at least four in the 4-year timeframe between 
Chandra and XMM-Newton observations.
One Chandra source disappears and a new one is detected by XMM-Newton in the 
ring.
Optical-UV colors of the Cartwheel ring are consistent with
a burst of star formation that is close to reaching its maximum, 
yielding a mean stellar age of about 40 Myr.
The inferred variability and age suggest that high mass X-ray binaries 
are the counterparts to the ULX sources.
The 3 companion galaxies have luminosities in the range 10$^{39-40}$ erg/s
consistent with expectations.
The hot gas of the Cartwheel galaxy is luminous and
abundant (a few $10^8$ $\mathrm{M_{\odot}}$) and is found both
in the outer ring, and in the inner part of the galaxy, behind
the shock wave front. We also detect gas in the group
with  $L_X \sim 10^{40}$ erg s$^{-1}$. }
{}


\authorrunning{Crivellari, Wolter \& Trinchieri}
\titlerunning{The XMM-Newton Cartwheel}

\maketitle

\section{Introduction}

The Cartwheel galaxy is a prototypical example of a ring galaxy,
located in a small and compact group of 4 members (SCG 0035-3357 -
Iovino 2002) at redshift z = 0.03. 
The group has a velocity dispersion of $\sim 400$ km s$^{-1}$ 
(Taylor \& Atherton 1984) and a radius of 60 kpc (R$_G$ = 1.7$^{\prime}$). 
However, three of its members (G1, G2,
and Cartwheel itself) are in a tight configuration of 
$< 1'$ radius, while G3, the most likely cause of the impact that
led to the formation of the Cartwheel (Higdon 1996), is situated
$\sim$ 3' ($\sim 100$ kpc) to the North.
The impact with G3 is likely to have transformed a normal spiral into
the Cartwheel galaxy today with its prominent 
elliptical rings, the outer
one with major axis $\sim80^{\prime\prime}$ ($45$ kpc) and
the inner one
$\sim 18^{\prime\prime}$
($10$ kpc). Optically emitting filaments, called ``spokes'',
connect the two rings.

\begin{figure}[!hutb]
%\centerline{\psfig{file=4cart-xmm-hst-new.ps,width=8cm,clip=} }
\centerline{\psfig{file=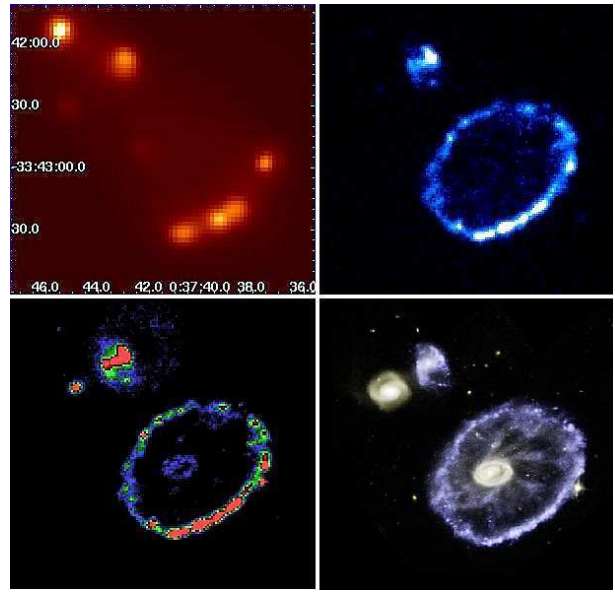,width=8cm,clip=} }
\centering

\caption{\footnotesize The Cartwheel group in: X-rays - XMM-Newton -
EPIC-MOS (top left). The bright X-ray source in the top left corner
is probably a background source; UV - XMM-Newton - OM (UVM2 filter; top right);
optical - XMM-Newton - OM (U filter; bottom left); optical - Hubble
Space Telescope (bottom right). The Cartwheel and galaxies G1 and G2 
are visible, while galaxy G3 is outside the area shown.
\label{cart-xmm}}

\end{figure}

The Cartwheel galaxy was observed at many wavelengths, from radio
(Higdon 1996), infrared (Marcum et al. 1992, Appleton \&
Struck-Marcell 1987), optical (Theys \& Spiegel 1976, Fosbury \&
Hawarden 1977, Struck et al. 1996), to ultraviolet (Gil de Paz et al.
2007) and X-ray bands. 
At every wavelength studied the emission from the outer ring was found
to be very high,
as exemplified in Fig.\ref{cart-xmm} where X-ray and optical-UV images
of the Cartwheel and its surroundings are shown.
The first observation in the X-ray band, obtained with \emph{ROSAT} in
1998, already revealed the prominence of the ring (Wolter, Trinchieri,
Iovino 1999). 
Later, in 2001, \emph{Chandra}
resolved a number of ULXs in the southern part of the ring, due to its 
excellent angular resolution ($\sim 0.5^{\prime\prime}$; Wolter
\& Trinchieri 2004, Gao et al. 2003).

We present \emph{XMM-Newton} data that we use to
study the ULXs in the ring,
the diffuse emission of the galaxy and the hot-gas group component.
In Sect. 2, we present the data; in Sect. 3, we discuss the analysis of
both the X-ray and optical/UV data; in Sects. 4, 5, 6, and 7, we
describe our results; and in Sect. 8, we summarize our findings.
We assume $H_0=75$ km s$^{-1}$ Mpc$^{-1}$ and z = 0.03, for which
D=122 Mpc and the angular scale is 0.56 kpc/arcsec.

\section{XMM-Newton data}

The Cartwheel galaxy was observed by \emph{XMM-Newton} on
2004 December 14-15  (Obs=[101]) for 36 ks and on 
2005 May 21-22 for 60 ksec (obs=[201]): in both sets of 
observations the EPIC-MOS and EPIC-pn instruments operated in FULL frame
mode with the THIN filter applied.

We have presented these X-ray datasets, which we used
to study the brightest ULX, source N.10 (Wolter, Trinchieri \& Colpi 2006).
We summarize the procedure used.
To exclude high flaring background periods, we used the XMM-Newton 
Science Analysis System (SAS - v6.5.0) to
extract the light curve of the photons with energy $\geq$ 10 keV, where
the signal was mostly due to particles, 
and we eliminated the periods of time in which the signal
was above 0.35 cts s$^{-1}$ and 1 cts s$^{-1}$ for
EPIC-MOS and EPIC-pn respectively. The final net exposure times were 29/24
ksec in the first observation and 50/42 ksec in the second one
(MOS/pn respectively) corresponding to 
a $\sim20 - 25 \%$ reduction in time.

The datasets from the two different epochs were both analyzed independently 
to study the
spectra and the variability of point sources, and  summed together to
improve the statistics for fainter sources and low surface brightness extended emission. 
For the spectral analysis, we used the EPIC-pn dataset. MOS1 and MOS2 were
used only when EPIC-pn data were unavailable.

To analyze the datasets we used, in addition to the {\sl xmm-sas} software, 
XSPEC (v11.3.1), ds9, ciao (v 3.4)
 \footnote{
http://xmm.esac.esa.int/external/xmm\_user\_support/documentation/sas\_usg/USG/;
http://heasarc.nasa.gov/docs/xanadu/xspec/manual/manual.html;\\
http://hea-www.harvard.edu/saord/ds9;
http://cxc.harvard.edu/ciao 
}.

\begin{table}[!ht]
\centering
\begin{tabular}{|c|c|c|c|c|}
  \hline
  & \scriptsize{\emph{Exp. Time}} & \scriptsize{$\lambda_{eff}$ } & \scriptsize{\emph{Conversion factor}} & \scriptsize{\emph{Zero point}}\\
  & ksec & \scriptsize{[\AA]} & \scriptsize{[erg cm$^{-2}$ s$^{-1}$ \AA$^{-1}$]}  &\\
\hline \hline
 \scriptsize{\emph{V}} & \scriptsize{8.8} & \scriptsize{5430} &  \scriptsize{2.49$\times10^{-16}$} &  \scriptsize{17.963}\\
 \scriptsize{\emph{B}} & \scriptsize{4.4} & \scriptsize{4500} & \scriptsize{1.29$\times10^{-16}$} &  \scriptsize{19.266}\\
 \scriptsize{\emph{U}} & \scriptsize{8.0} & \scriptsize{3440} & \scriptsize{1.94$\times10^{-16}$} &  \scriptsize{18.259}\\
 \scriptsize{\emph{UVW1}} & \scriptsize{8.8} & \scriptsize{2910} & \scriptsize{4.76$\times10^{-16}$} &  \scriptsize{17.204}\\
 \scriptsize{\emph{UVM2}}&  \scriptsize{8.8} & \scriptsize{2310} & \scriptsize{2.20$\times10^{-15}$} &  \scriptsize{15.772}\\
 \scriptsize{\emph{UVW2}} & \scriptsize{14.4} & \scriptsize{2120} & \scriptsize{5.71$\times10^{-15}$} &  \scriptsize{14.867}\\
 \hline
\end{tabular}
\centering
%\begin{minipage}{0.45\textwidth}
\caption{Effective wavelengths and conversion factors (from count rate to flux) of OM filters. \label{OM}
}
%\end{minipage}
\end{table}

We also analyzed the optical/UV
data obtained with the Optical Monitor onboard XMM-Newton:
the first set of observations provides data in the  V, B, 
and U filters, 
while for the second set, UV data are obtained with UVW1, UVM2, and UVW2 
filters.  
Exposure times, central wavelength of the filter, conversion factors, and 
zero-point magnitudes are listed in Table~\ref{OM}\footnote {from http://xmm.vilspa.esa.es/external/xmm\_user\_support/documentation/uhb/node75.html}.

\section{Data analysis}

\subsection{EPIC data}

We first studied the point sources of the Cartwheel galaxy. We created
0.3 - 7 keV images for EPIC-MOS and EPIC-pn separately from both sets
of observations 
(see Fig.~\ref{cart-xmm}), and applied the standard detection algorithm in the
area around the Cartwheel simultaneously to all datasets from the same epoch. 
We detected 6 and 7 sources in the area covered by the disk Cartwheel 
for the dataset of 
first and second observations, respectively, for a total of 8 sources.  
We also detected 
G1 and G2 and another source in the vicinity, which is most likely 
to be in the background, in 
both observations.  All detected sources are listed in Table~\ref{srclist}. 
To estimate the source strength, we used the final positions given by the 
detection procedure, but 
recomputed the net counts in circles of radius 10$^{\prime\prime}$ 
(with the exception of source XMM4, in which r = 8$^{\prime\prime}$
to avoid a CCD gap in the EPIC-pn 
image in obs. [101]), which is
smaller than normal but avoids inclusion of nearby sources and
a too-large fraction of the surrounding ring-emission.
The background for sources XMM9, XMM10, and XMM11, outside the 
Carthwheel's ring, is computed in a circle of radius 35$^{\prime\prime}$ 
outside the ring and the group area, devoid of sources.
For sources in the Cartwheel,  we compute the ``galaxy'' background
with two circles of radius 12$^{\prime\prime}$ and 10$^{\prime\prime}$
including the ring and part of the region inside it but not 
the detected sources.  
Since we also intend to compare with Chandra data, 
the sources are arranged in the order of their Chandra counterparts (from 
Wolter \& Trinchieri, 2004), which  are listed in Table~\ref{srclist}.
Source XMM2=N10 was already 
presented and discussed in Wolter et al. (2006). It is listed in 
Table~\ref{srclist} for completeness, but is not discussed further, 
except when relevant to the other sources.  
Using a simple power-law spectrum ($\Gamma = 2.2$ and
n$_{H} = 1.9 \times 10^{21}$ cm$^{-2}$, which was derived from the sum of detected 
sources in Chandra, see Wolter \& Trinchieri 2004) and the distance of 122 Mpc,
we also compute fluxes and the intrinsic luminosity of each source, 
which we list in  Table~\ref{srclist}. 
We confirm the high luminosity of each XMM-Newton detections 
($\geq 10^{39}$ erg/s), as already noted from the Chandra results, 
that places each of them  in the  range of ULX sources.

We performed spectral analysis for the brightest sources (with about $\geq 100$ net
counts). We binned the data so that each final bin has a significance of at
least 2$\sigma$ in the net data, and we fitted the spectra with a power-law 
model including low energy
absorption caused by an intervening column with free N$_H$.

\begin{table*}[!ht]
\centering
\begin{tabular}{|c|cc|c|c|c|c|c|c|}
  \hline
XMM name  & \scriptsize{RA} & \scriptsize{Dec} & \multicolumn{2}{|c|}{\scriptsize{\emph{pn Net counts(0.3-7 keV) }} } & \scriptsize{\emph{Chandra name}}& \multicolumn{2}{|c|}{\scriptsize{\emph{log L$_X$ erg/s(0.5-2 keV)}}} & \scriptsize{\emph{Detected} } \\
  & \multicolumn{2}{|c|}{(J2000)} & [101] & [201] &  & [101] & [201] &  \\
\hline \hline
%XMM1 N6       1=no; 2=no
%XMM2 N7+N9    1=si; 2=si
%XMM3 N10      1=si; 2=no
%XMM4 N11      1=si; 2=si
%XMM5 N12      1=si', 2=si' raggio piu' piccolo x gap in [pn101]
%XMM6 N13+N14  1=no; 2=si'
%XMM7 N16+N17  1=si; 2=si'
%XMM8 N21      1=no; 2=si'
%XMM9  NEW    1=si; 2=si'
%XMM10 G1       1=si; 2=si
%also G2  1=si; 2=si --> in G2.outsave
%XMM11 CXOJ003745.6-334151 (aveva 233 Ch counts) 1=si; 2=si'   89.534    10.296   344.407    19.334  

%XMM1  & 00:37:42.3 & -33:43:03.4 & $<$ 5.95 & $<$7.95 & \scriptsize{\emph{\textbf{N6} }} && & 1:no; 2:no \\
XMM1  & 00:37:40.8 & -33:43:32.5 & 110.94$\pm$ 12.94& 82.36$\pm$13.14 & \scriptsize{\emph{\textbf{N7+9} }}  &40.06 &39.69 & 1:yes; 2:yes \\
XMM2  & 00:37:39.4 & -33:43:23.2 & 209.09$\pm$16.35 & 184.43$\pm$16.63 & \scriptsize{\emph{\textbf{N10} }}  & 40.33& 40.04& 1:yes; 2:no \\
XMM3  & 00:37:39.0 & -33:42:49.3 & 12.80$\pm$8.24 & 16.77$\pm$10.22 & \scriptsize{\emph{\textbf{N11} }}  &39.12&39.00 & 1:yes; 2:yes \\
XMM4$^{*}$  & 00:37:39.2 & -33:42:32.76 & 4.68$\pm$6.18 & 19.78$\pm$8.45 & \scriptsize{\emph{\textbf{N12} }}  &38.68&39.07 & 1:yes; 2:yes \\
XMM5  & 00:37:39.0 & -33:43:19.2 & 153.33$\pm$13.89 & 202.89$\pm$17.15 & \scriptsize{\emph{\textbf{N13+14} }}  &40.20& 40.08& 1:no; 2:yes \\
XMM6  & 00:37:37.5 & -33:42:56.9 & 47.94$\pm$10.23 & 96.36$\pm$13.66 & \scriptsize{\emph{\textbf{N16+17} }}  &39.69& 39.76& 1:yes; 2:yes \\
XMM7  & 00:37:40.9 & -33:42:33.5 & $<$ 7.40 & 31.89$\pm$11.09 & \scriptsize{\emph{\textbf{N21} }}  &$<$ 38.38& 39.28& 1: no; 2:yes \\
XMM8  & 00:37:42.2 & -33:42:51.5 & 13.94$\pm$8.40 & 30.89$\pm$11.05 & \scriptsize{\emph{\textbf{none} }} &39.16&39.26 & 1:yes; 2:yes \\
XMM9  & 00:37:43.1 & -33:42:07.9 & 132.42$\pm$12.21 & 199.00$\pm$15.09 & \scriptsize{\emph{\textbf{G1} }} &40.13& 40.07& 1:yes; 2:yes \\
XMM10  & 00:37:45.1 & -33:42:28.4 & 33.53$\pm$7.07 & 80.41$\pm$10.48 & \scriptsize{\emph{\textbf{G2} }}  &39.54&39.68 & 1:yes; 2:yes \\
XMM11  & 00:37:45.5 & -33:41:52.8 & 89.53$\pm$10.30 & 344.41$\pm$19.33 & \scriptsize{\emph{\scriptsize{CXOJ003745.6-334151} }}  &39.96&40.55 & 1:yes; 2:yes \\

 \hline
\end{tabular}
\centering
%\begin{minipage}{0.4\textwidth}
\caption{Names, positions, and counts. Background includes region of the Cartwheel ring and inside, but not the detected sources. $^{*}$: radius=8'' close to pn gap in [101]
\label{srclist}
}
%\end{minipage}
\end{table*}

\begin{figure}[!h]
\begin{minipage}[l]{0.45\textwidth}
\psfig{file=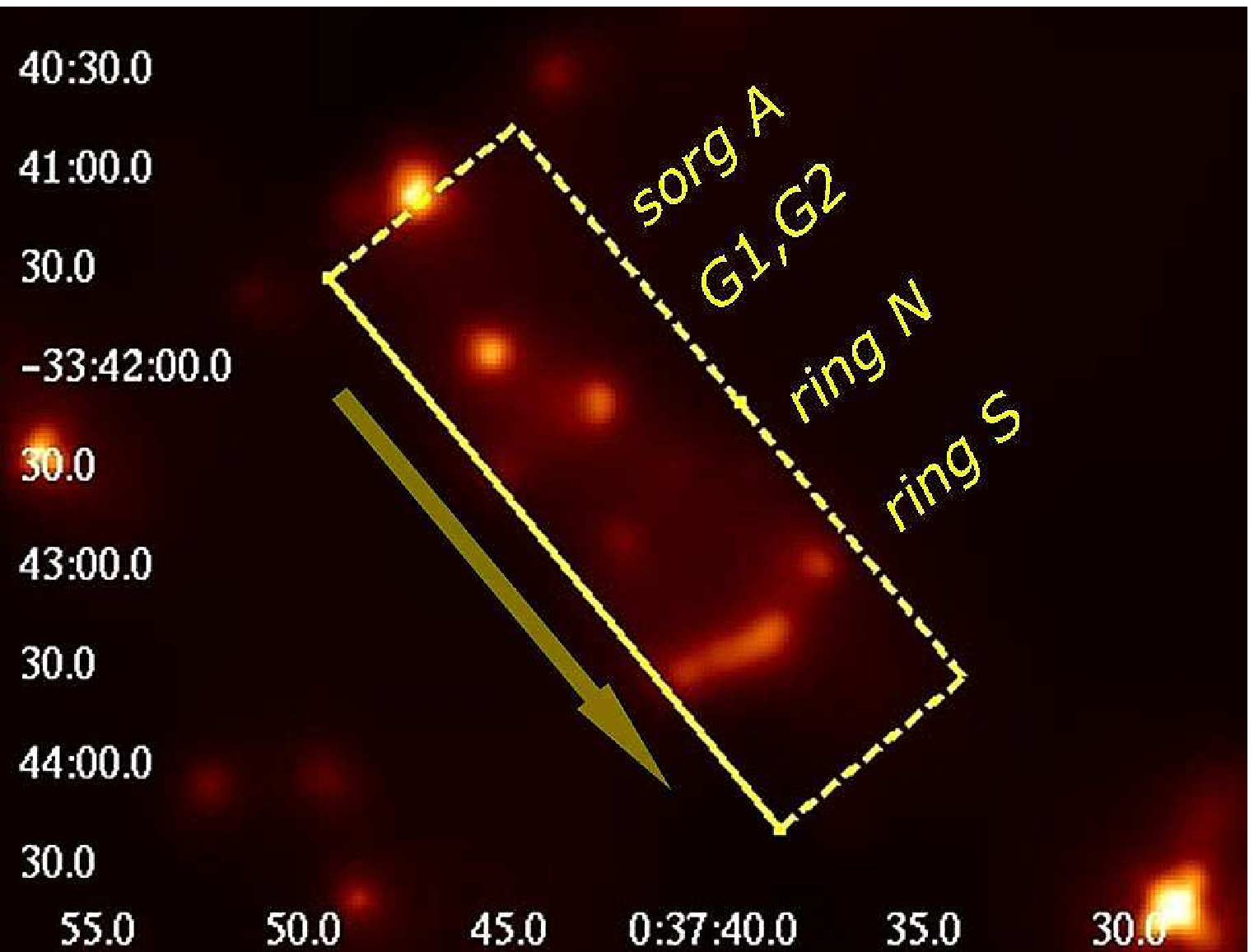,width=7cm,clip=}
\end{minipage}
\begin{minipage}[r]{0.45\textwidth}
\psfig{file=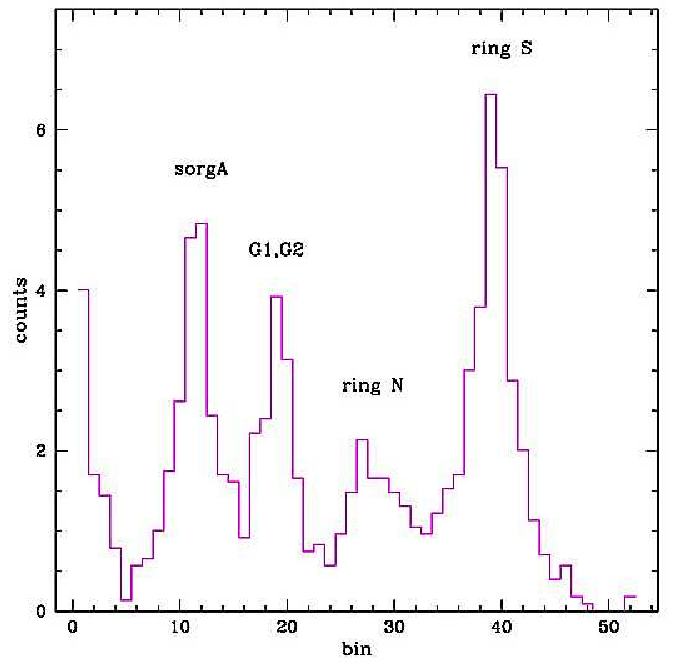,width=6cm,clip=}
\end{minipage}
\centering
%\begin{minipage}{0.40\textwidth}
\caption{Left: projection region (thickness = 72$^{\prime\prime}$, length = 215$^{\prime\prime}$); the
background is derived from a box, parallel to this one, displaced to
the SE. Right: projection of the background-subtracted counts
(in NE-SW direction). \label{projection}}
%\end{minipage}
\end{figure}

We also studied the unresolved component in the galaxy to determine its
spectral characteristics and luminosity. We first made a
projection of the net counts in the region shown in Fig.\ref{projection}.
We detected a positive signal inside
the Cartwheel ring and in the region between the Cartwheel and its two
nearby companion galaxies, which appears to represent a contribution
from both the body of the Cartwheel and from the group.

\begin{figure}[!htb]
%\centerline{\psfig{file=gas+bg-new.ps,width=8cm,clip=} }
\centerline{\psfig{file=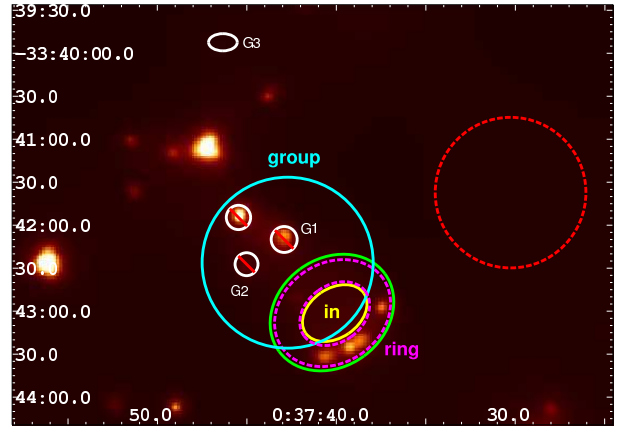,width=8cm,clip=} }
\centering
%\begin{minipage}{0.6\textwidth}
\caption{The 0.3 - 2 keV MOS1+MOS2 image, with the regions used to extract 
the spectra and the background region overplotted %in different colors
as labeled. The background region is the dashed circle.
All detected sources are excluded in the calculation, but not all the
corresponding excluded regions are shown, for clarity.  \label{gas}}
%\end{minipage}
\end{figure}

To study the spectral characteristics of the unresolved emission in the 
Cartwheel, we considered both a region corresponding to the ring (``ring''),
and one inside it  (``in''), as shown in Fig.~\ref{gas}. 
All detected sources were masked out.  We then summed the 
resulting spectra from the two observations and combined the response matrixes 
(weighted by the relative exposure times).
With the same procedure, we also obtained the  spectral data
for the companion galaxies, G1 and G2 (G3 has an insufficient number of counts), 
and for a region between them, but outside the Cartwheel's ring, 
where we detected emission from the group. 

We used a power-law model to represent an unresolved binary 
component, while the hot plasma component was described by a \emph{mekal} 
model with abundances fixed at 0.5$\times$ solar, 
since the optical metallicity is low (Fosbury \& Hawarden, 1977).
The choice of abundances however had no influence on the final 
results for kT.

The spectral results are discussed in Sects. 4, 5, and 6 for the point sources,
the companion galaxy, and the diffuse emission, respectively.

\subsection{OM data}

OM data were processed in {\sl XMM-SAS}  using the 
standard parameters described in the XMM-Newton Science Analysis 
System Users' Guide.

For the brightest sources, we calculated the net
counts in a region of radius 2$^{\prime\prime}$ for the optical filters, and
3$^{\prime\prime}$ for the UV ones, centered on the pixel with the highest signal  
consistent with the \emph{Chandra} position.
A mean background was evaluated by computing the average surface brightness 
within different circular regions
outside but still close to the galaxy.
Given the presence of ghost images, which were more prominent in some filters, 
we chose different background regions appropriately
for each source to minimize the contamination from the ghost images.
With the same procedure we also computed net counts from the region 
corresponding to the
brightest Mid-IR hot spot (an HII region; see Charmandaris et
al. 1999) %(3") 
and from the galaxies G1 and G2. % (11").

We then used the conversion factors listed
in Table~\ref{OM} to  derive fluxes and Vega magnitudes of these sources. 
Our results are described in  Sect. 7.

\section{Point sources and variability}

\begin{table*}[!ht]
\centering
\begin{tabular}{|c|c|cc|cc|cc|cc|}
 \hline
 & 
 \scriptsize{$\Gamma$} & \multicolumn{2}{|c|}{\scriptsize{$n_H$ } }& \multicolumn{2}{|c|}{\scriptsize{$\chi^{2}$}} &
 \multicolumn{2}{|c|}{\scriptsize{\emph{Flux}} } &
 \multicolumn{2}{|c|}{\scriptsize{\emph{Lum}} }\\
 & &
 \multicolumn{2}{|c|}{\scriptsize{[cm$^{-2}$]$\times10^{21}$} }& \multicolumn{2}{|c|}{} &
 \multicolumn{2}{|c|}{\scriptsize{($2-10$ keV) [erg cm$^{-2}$ s$^{-1}$]$\times10^{-14}$} } &
 \multicolumn{2}{|c|}{\scriptsize{($2-10$ keV) [erg s$^{-1}$]$\times10^{40}$} }\\
 & & \scriptsize{[101]} & \scriptsize{[201]} & \scriptsize{[101]} & \scriptsize{[201]} & \scriptsize{[101]} & \scriptsize{[201]} & \scriptsize{[101]} & \scriptsize{[201]} \\
 \hline
 \hline
 \scriptsize{\emph{\textbf{N7+N9} }} &  \scriptsize{$\Gamma=1.9$} & \scriptsize{$1.7\pm0.6$} & \scriptsize{$1.8\pm0.5$} &
 \scriptsize{$0.35$} & \scriptsize{$0.34$} & \scriptsize{1.59[1.22-1.94]} & \scriptsize{0.89[0.60-1.20]} & \scriptsize{2.94[2.32-3.52]} & \scriptsize{1.62[1.14-1.97]} \\
% \scriptsize{\emph{\textbf{N10}} (*)} &  \scriptsize{$233$} & \scriptsize{$218$} & \scriptsize{$8.9$} & \scriptsize{$4.7$} & \scriptsize{$\Gamma=1.6$} & \scriptsize{$2.3$} & \scriptsize{$2.3$} & & & \scriptsize{$5.0$} & \scriptsize{$2.3$} \\
 \scriptsize{\emph{\textbf{N13+N14}}} &  \scriptsize{$\Gamma=2.1/1.4$} & \scriptsize{$2.3\pm0.7$}& \scriptsize{$0.6\pm0.3$}&
 \scriptsize{$0.97$} & \scriptsize{$0.92$} & \scriptsize{1.55[1.21-2.02]}&  \scriptsize{1.51[1.09-1.80]} & \scriptsize{2.81[2.34-3.76]} & \scriptsize{2.68[2.11-3.34]} \\
 \scriptsize{\emph{\textbf{N16+N17}}} &  \scriptsize{$\Gamma=2.2$} & \scriptsize{$2.0\pm0.6$} & \scriptsize{$2.0\pm0.6$} &
 \scriptsize{$0.78$} & \scriptsize{$1.2$} & \scriptsize{0.56[0.34-0.85]} & \scriptsize{0.71[0.49-0.90]}& \scriptsize{1.04[0.67-1.38]}& \scriptsize{1.30[0.89-1.65]} \\
\hline
\end{tabular}
\centering
%\begin{minipage}{1\textwidth}
\caption{Spectra, fluxes, and luminosities of the three brightest ULX in the
ring. %(*) The values of N10 are from Wolter \& Trinchieri (2004) and Wolter et al. (2006).
  \label{tab1} }
%\end{minipage}
\end{table*}

\begin{figure*}[htb]
%\centerline{\psfig{file=anello_ok2.ps,width=16.5cm,clip=} }
%\centerline{\psfig{file=fig4-cart.ps,width=16.5cm,clip=} }
%\psfig{file=fig4-cart.ps,width=16.5cm,clip=} 
%\psfig{file=fignuova-fin.ps,width=16.5cm,clip=} 
\psfig{file=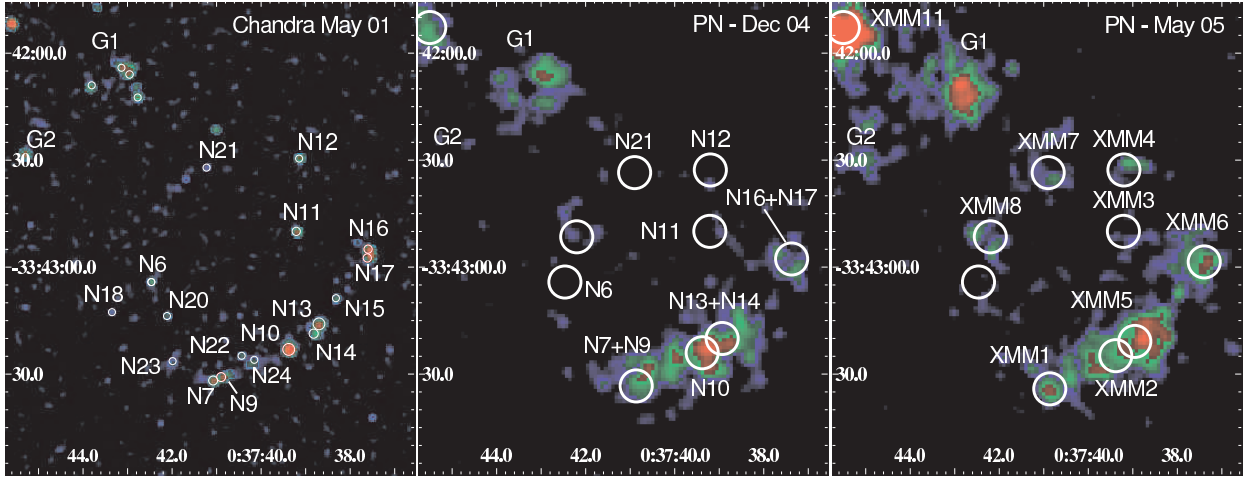,width=16.5cm,clip=} 
\centering
%\begin{minipage}{0.92\textwidth}
\caption{The three panels represent a zoomed image of the ring on the same scale. Left:
0.3 - 2 keV \emph{Chandra} image. %Only the sources that correspond to \emph{XMM-Newton} detections are indicated.
Middle and right: 0.3 - 2 keV EPIC-pn images, smoothed with a 3$^{\prime\prime}$ Gaussian, of the two epochs respectively.
Source names of the detected sources are indicated using
\emph{Chandra} names in the middle panel and using \emph{XMM-Newton}
names in the right panel. \label{3ring}}
%\end{minipage}
\end{figure*}

Seven sources were detected in the Cartwheel ring in the XMM-Newton data.  
One additional source was found in the region inside the ring (XMM3=N11) 
and 1 outside (XMM11).  G1 and G2 were also clearly detected.

A zoomed image of the southern portion of the ring is
shown in Fig.\ref{3ring} and compared with the previous Chandra image
of the same region. 
Because of the lower XMM-Newton
resolution (FWHM $\sim 6^{\prime\prime}$), many XMM-Newton sources
correspond to two Chandra ones. 

Five sources (XMM1, XMM3, XMM4, XMM6, XMM8) plus the three sources clearly 
not associated with the Cartwheel (XMM9, XMM10, and XMM11) 
are detected in both sets of observations.
Source XMM2 is found only in the first 
dataset, while sources XMM5 and  XMM7 were detected in the second set.  
All sources except XMM8 were previously detected by Chandra, as clearly indicated by the associations listed in Tab.~\ref{srclist}.

\begin{figure}[!htb]
%\centerline{\psfig{file=grafico_ringsources_XMM_chandra_log.ps,width=8cm,clip=} }
%\centerline{\psfig{file=grafico_new.ps,width=8cm,clip=} }
\centerline{\psfig{file=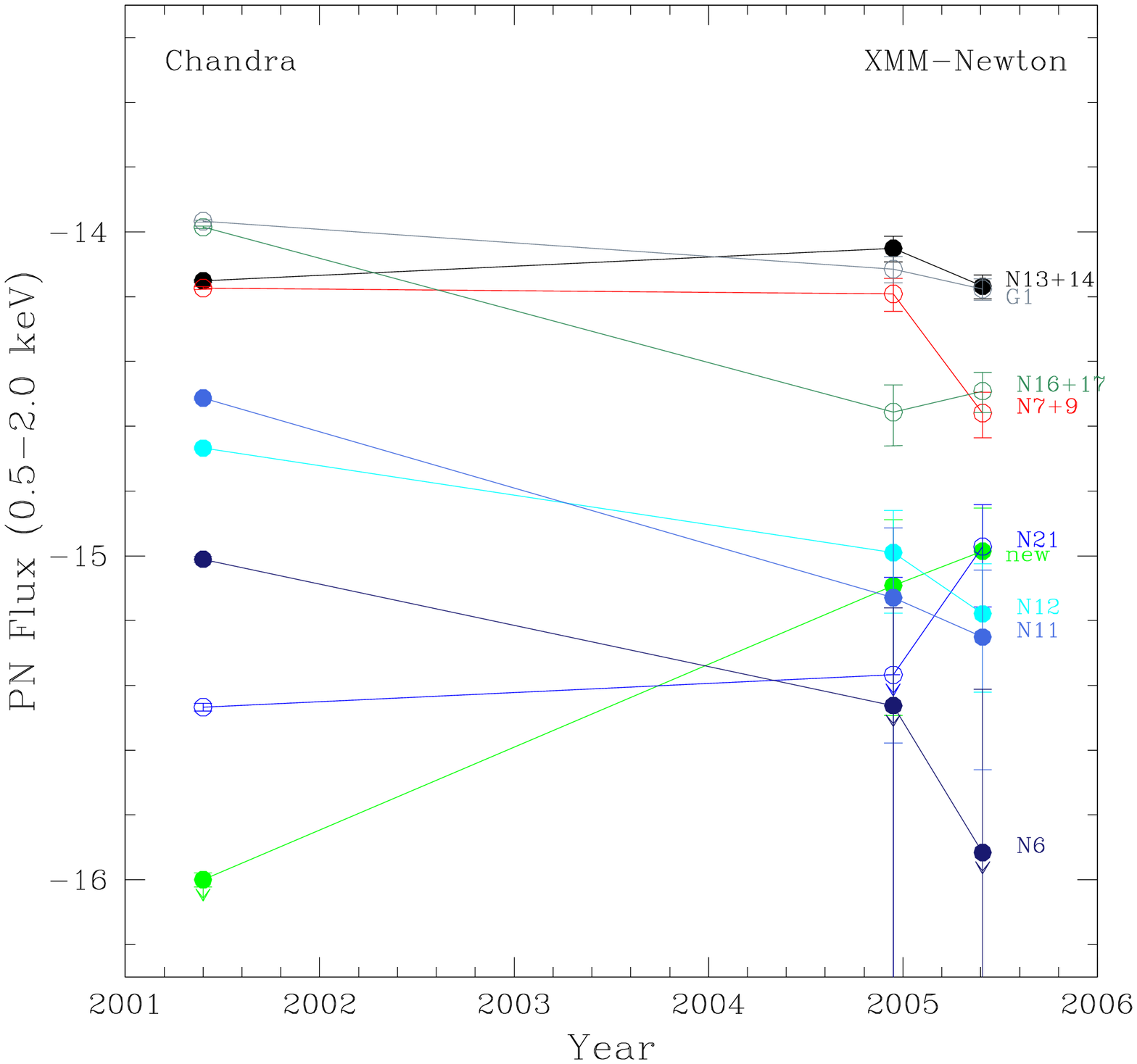,width=8cm,clip=} }
\centering
%\begin{minipage}{0.45\textwidth}
\caption{Comparison of fluxes in the 4-year time-frame covered by
\emph{Chandra} and \emph{XMM-Newton}. Plotted errorbars
include only statistical errors. The same spectral shape is used for
all observations, see text for details. Chandra values are from Wolter
\& Trinchieri (2004). \label{lightcurve}}
%\end{minipage}
\end{figure}

We compare XMM-Newton fluxes derived in Sect. 3.1 with Chandra fluxes from 2001.
The resulting light curves are plotted in Fig.~\ref{lightcurve}.
 Given the different spatial resolutions, we summed the contribution from
two Chandra sources when more than one was included in the larger XMM-Newton
PSF (as indicated by their names). 
We also estimated the values relative to source N.6, the brightest
source detected in the Chandra image and formally undetected in the XMM-Newton datasets.  For source XMM8 previously undetected, we assumed a flux at the Chandra epoch equivalent to the lowest detected source in the ring or vicinity as a reasonable upper limit to its flux. 

We observed fluctuations in the flux of the sources in the two observations,
even if they were not of large amplitude. 
Sources N.7+9 and N.13+14 show a decrease in count rate, while in N.21 there is an increase.  
Other sources show fluctuations either way, but they should be considered constant within the uncertainties. 

We detected evidence of variability on timescales of a few years in several cases: 
N.16+17, N.11, N.12, N.21, and N.6 varied significantly in brightness 
between the Chandra and the XMM-Newton first observation, and XMM8 has 
significantly increased in brightness since 2001.

For the three brightest sources detected with $>100$ EPIC-pn counts, we 
attempted a spectral fit.
 For consistency with Chandra results, we used a background level measured
outside the Cartwheel disk, and therefore we needed to use a complex model 
to account for different components, since the ring is likely to contribute 
significantly. % amount in the extraction region.
We  used the model derived from the Chandra data to parameterize the ring emission (\emph{mekal} with  $\mathrm{kT}=0.2$ keV,
to account for a hot gas component, and 
\emph{powerlaw}  with $\Gamma = 2.3$ for the unresolved
binary population, with the 
relative normalizations fixed to the Chandra values, Wolter \& Trinchieri 2004), and rescaled it  to the area subtended by the source region.
The source spectrum was parameterized by a power-law model with low 
energy absorption. 
In Table~\ref{tab1}, we list the photon index, and the fitted 
absorption, which is always much greater than Galactic,
fluxes and luminosities for the ULX component only, i.e. 
cleaned of the contribution from the ring parametrized as
discussed above.
%above the ``ring'' contribution from the two components defined above.

\begin{figure}[!htb]
%\centerline{\psfig{file=N13N14spectra.ps,width=8.2cm,clip=} }
\centerline{\psfig{file=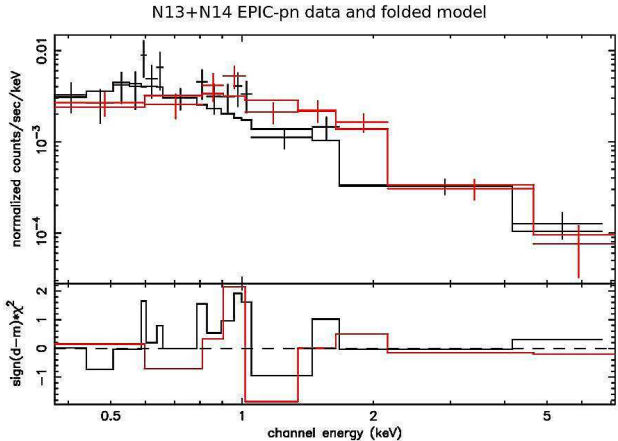,width=8.2cm,clip=} }
\centering
%\begin{minipage}{0.65\textwidth}
\caption{Spectra of source N13+14; in red: data from 1st observation - Dec 2004 fitted with $\Gamma=2.1$ and n$_H=2.3\times10^{21}$ cm$^{-2}$;  in black: data from 2nd observation - May 2005 fitted with $\Gamma=1.4$ and n$_H=6.5\times10^{20}$ cm$^{-2}$.   \label{N13N14spectra}}
%\end{minipage}
\end{figure}

Source N13+14 deserves closer scrutiny. For example it is possible 
that the presence of source N.10 at $\sim 10^{\prime\prime}$ from
N13+14 prevents the detection of this latter source in the first observation,
since it contributes only a small fraction of the emission and is
therefore difficult to separate from the much brighter source N.10.
In contrast, in obs. [201], N.10 has dimmed (Wolter et al. 2006)
and was not formally detected, while N13+14 was.
We also note that in the
first XMM-Newton observation the source appears brighter than in the
Chandra data, which again could be due to contamination from the wings of
the brighter source.  The spectral data of the 
two XMM-Newton observations indicate a mild count
rate variability and different best-fit values, 
as shown in Fig.~\ref{N13N14spectra}: the photon
index varies from 2.1 [1.66 - 2.66] to 1.4 [1.00 - 1.97], 
and the absorption coefficient also 
decreases in the second observation (from $2.3[1.4-3.4]\times 10^{21}$ to
$6.5[2.9-11.4]\times 10^{20}$ cm$^{-2}$).\footnote{Note that the flux reported 
in Fig.~\ref{lightcurve} is derived by assuming the same spectral shape 
with $\Gamma=2.2$}
Unfortunately, the statistical significance of the two observations is
low. However, it appears that the spectral variation is in a sense opposite to
that expected from strong contamination from source N.10: the
spectral shape of this latter is $\Gamma=1.6$ and should act in an opposite
sense to the spectral change observed.

All point sources exhibit some type of variability either on 5-month or 4-year
timescales.  A few sources vary by up to a factor of $\geq$ 10 in flux.
This is a clear indication that the X-ray source is an individual
accreting object. The statistics of the individual detections do not
allow a more precise comparison with e.g. the Galactic black hole
spectral - luminosity plane. Variability on timescales of
months to years 
is also observed in other galaxies. In the most studied of them, the
Antennae (Zezas et al. 2006), most ULXs exhibit long-term variability,
of factors $\sim$ 2-6. Source 39 exhibits a strong decline to very faint fluxes
($\Delta f_X > 10$), which could be similar to the behavior of our
sources XMM8 and N6.

\section{Companion galaxies}

\begin{table*}[t]
 \begin{tabular}{|c|c|c|c|c|c|c|c|}
 \hline
  & & \scriptsize{\emph{\textbf{``Gas''}} } & \multicolumn{2}{|c|}{\scriptsize{\emph{\textbf{``Stellar'' component}}}} & & & \scriptsize{\emph{\textbf{$\chi^2$ (dof)}} }    \\
 & \scriptsize{\emph{EPIC-pn Count rate}} &
\scriptsize{$L_{x}$($0.5-2 \mathrm{keV}$)/ kT} &
\scriptsize{$L_{x}$($0.5-2 \mathrm{keV}$)}
 & \scriptsize{$L_{x}$($2-10 \mathrm{keV}$)}  
& \scriptsize{$\Gamma$} & \scriptsize{\emph{\textbf{$n_H$}} } &  \scriptsize{}\\
 & \scriptsize{($0.3-7$keV)[s$^{-1}$]} &
\scriptsize{[erg s$^{-1}$]/kT} &
\scriptsize{[erg s$^{-1}$]}
 & \scriptsize{[erg s$^{-1}$]}  
& & \scriptsize{[cm$^{-2}$] $\times10^{20}$ } & \\
 \hline\hline
\scriptsize{\emph{\textbf{G1}}} & \scriptsize{$(6.4\pm0.3)\times10^{-3}$} & \scriptsize{$(8.6\pm0.9)\times10^{39}$/} & \scriptsize{$(2.1\pm0.3)\times10^{40}$} & \scriptsize{$(1.9\pm0.2)\times10^{40}$} &
\scriptsize{1.86[1.26 � 2.49]} &
\scriptsize{17.[13.  � 82] }  & \scriptsize{0.94(24)}   \\
\scriptsize{} &  & \scriptsize{0.22[0.16  � 0.28]} &  & & & &  \\
\scriptsize{\emph{\textbf{G1}}} & \scriptsize{\it idem } & \scriptsize{--} & \scriptsize{$(2.4\pm0.2)\times10^{40}$} & \scriptsize{$(1.5\pm0.2)\times10^{40}$} &
\scriptsize{2.15[1.8 � 2.5]} &
\scriptsize{9.[3.  � 16.] }  & \scriptsize{1.09(26)}    \\
\scriptsize{\emph{\textbf{G2}}} & \scriptsize{$(2.9\pm0.2)\times10^{-3}$} & \scriptsize{--} & \scriptsize{$(8.1\pm2.1)\times10^{39}$} & \scriptsize{$(5.4\pm1.3)\times10^{39}$} &
\scriptsize{2.35[2.06 � 2.98]} &
\scriptsize{8.9[3.8  � 17.8] }  & \scriptsize{1.12(16)}   \\
\scriptsize{\emph{\textbf{G3}}} & \scriptsize{$^{*}(2.6\pm0.9)\times10^{-4}$} & \scriptsize{--} & \scriptsize{$(3.0\pm1.0)\times10^{39}$} & \scriptsize{$(1.9\pm0.7)\times10^{39}$} &
\scriptsize{2.15} &
\scriptsize{9.0} &  \\
 \hline
 \end{tabular}
\centering
%\begin{minipage}{0.83\textwidth}
\caption{Count rates and luminosities of Cartwheel companion
galaxies. Results for G1 are given for both 2 model components (1st line)
and single power law (2nd line). {$^{*}$ Count rate for G3 is from MOS1+MOS2} \label{G1G2G3}}
%\end{minipage}
\end{table*}

We extracted the spectra of the companion galaxies from the two
EPIC-pn datasets %, which have the larger statistics, 
and we coadded the data since we did not expect significant time
 variability (see also Fig.~\ref{lightcurve}).
We attempted to represent the spectrum with a
\emph{powerlaw} model plus a \emph{mekal} model, to account for both
the stellar component and the gas contribution, when present.
In Table~\ref{G1G2G3}, we summarize the count rates and luminosities of these galaxies.

The spectrum of G1 is well fitted by a single component, an absorbed 
\emph{powerlaw}, with slope $\Gamma = 2.15$ and
n$_H=1.4 \times 10^{21}$ cm$^{-2}$, which implies a total $L_X(0.5-10) =
3.9 \times 10^{40}$ erg s$^{-1}$.
Similar parameters characterize the G2 spectrum: 
$\Gamma = 2.4$ and n$_H=9. \times 10^{20}$ cm$^{-2}$, for a 
total $L_X(0.5-10) = 1.4 \times 10^{40}$ erg s$^{-1}$, with 
larger uncertainties because of the smaller number statistics. 
In both cases, as for the Cartwheel galaxy itself,
 the line-of-sight Galactic value of n$_H=1.94\times 10^{20}$ cm$^{-2}$ 
(Dickey \& Lockman 1990) is about a factor of 10 lower, which is indicative 
of absorption within the galaxies.

While formally not required by the fitting procedure, the addition of a
\emph{mekal} component to the spectral fit of the G1 data
improves the distribution of the residuals, which show 
an excess at 0.6-0.8 keV above the single power law model. 
This component has a temperature of kT=0.22 keV and an unabsorbed
luminosity of L$_X^{0.5-2} = 8.6 \times 10^{39}$ erg/s. (see Table~\ref{G1G2G3}).

The position of G3 coincides with a bad column in EPIC-pn in both 
observations.  Therefore we used
the MOS1 plus MOS2 data (a total of 51 net counts), which allow us only
to measure the normalization from which we derived the luminosities. 
By assuming the same power-law spectrum that describes G1,
we measured a L$_X^{0.5-2} = 3 \times 10^{39}$ erg/s.

\begin{figure}[!h]
%\centerline{\psfig{file=G1_chandra.ps,width=6cm,clip=} }
\centerline{\psfig{file=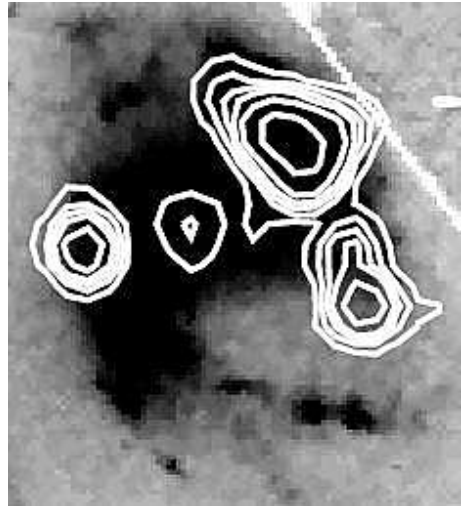,width=6cm,clip=} }
\centering
%\begin{minipage}{0.52\textwidth}
\caption{The X-ray contours in the 0.3-7 keV Chandra image overplotted on the optical HST image (adapted from Wolter \& Trinchieri 2004). \label{G1_chandra}}
%\end{minipage}
\end{figure}

Higdon (1996) evaluated intrinsic %({\bf corrected? spiegare meglio}) 
B luminosities of G1 and G2:
$L_B$(G1)$=6\times 10^9 L_{\odot}$ and $L_B$(G2)$=4.8\times 10^9
L_{\odot}$. There is no comparable measuremente in the literature for G3.
Using the relation from Kim \& Fabbiano (2004)
we expect that Low Mass X-ray Binaries (LMXBs) contribute
L$_X^{G1}(0.3-8 keV) = 5.4\times 10^{39}$ erg/s; L$_X^{G2}(0.3-8 keV) = 4.\times10^{39}$ erg/s 
respectively. The observed L$_X$ is a factor of 7 and 4 higher  
in G1 and G2, respectively. % {\bf lX(0.3-8 keV) viene circa 4 e40 per G1; e 1.7e40 per G2}. 
At the \emph{XMM-Newton} spatial resolution we cannot exclude
the presence of a nuclear source in G1, however we know from the 
\emph{Chandra} data that the emission is caused by 
a number of mostly off-center sources (see Fig.~\ref{G1_chandra}).
Since the total G1 luminosity in the 2-10 keV band has not varied 
between Chandra and  \emph{XMM-Newton} observations, 
we expect that the same sources contribute to the overall emission 
detected with XMM-Newton.  
Therefore, the relatively high $L_X$ in the companion galaxies,
and in particular G1, is probably related to
the interaction between the Sc galaxy G1 and the S0 companion G2, as
also indicated by HI observations (Higdon 1996), 
which has resulted in  morphological distortions, faint tidal tails,
a higher SFR, and possibly a high number of HMXBs.
Galaxy G3, now considered the ``intruder'' (Higdon 1996), has been
poorly sampled at different wavelengths. 
%despite having being labeled as the ``intruder'', has very
%sparse information at different wavelengths. 
Nevertheless, if it has the
same corrected L$_B$ as the other two companions, as appears to
be the case according to their similar 
observed optical and IR magnitudes, its X-ray luminosity
is in line with expectations from Kim \& Fabbiano (2004).

\section{Diffuse emission}

\begin{table*}[!ht]
\centering
\begin{tabular}{|c|ccc|c|c|c|c|c|}
  \hline
   & \multicolumn{3}{|c|}{\scriptsize{ \emph{Gas}}} & \scriptsize{$n_H$} & \scriptsize{$\chi^2$ (dof)}  & \scriptsize{\emph{n}} & \scriptsize{\emph{V}} & \scriptsize{\emph{M}} \\
 \hline
 & \scriptsize{\emph{kT}} & \scriptsize{\emph{pn Count rate }} & \scriptsize{\emph{Flux}}  & & &&& \\
 & & \scriptsize{(0.3-2 keV)} & \scriptsize{(0.5-2 keV)} & & & \scriptsize{[cm$^{-3}$]} & \scriptsize{[cm$^{3}$]}  & \scriptsize{[M$_{\odot}$]} \\
 & & \scriptsize{$\times 10^{-3}$} & \scriptsize{$\times 10^{-14}$} &  \scriptsize{$\times 10^{21}$} & &&&\\
 \hline
 \hline
 \scriptsize{\emph{Ring}} & \scriptsize{0.20 [0.16-0.22]} & \scriptsize{$10.09 \pm 0.43$} &   \scriptsize{$1.18$} & \scriptsize{$0.23$} &   \scriptsize{$0.73$(24)}  & \scriptsize{2.4$\times 10^{-3}$} &  \scriptsize{2.03$\times 10^{68}$} &  \scriptsize{4.7$\times 10^{8}$} \\
 \scriptsize{\emph{``In''}} & \scriptsize{0.16 [0.12-0.21]} & \scriptsize{$7.02 \pm 0.37$} &   \scriptsize{$1.03$} & \scriptsize{$2.9 \pm 0.9$} &   \scriptsize{$0.99$ (26)}  & \scriptsize{2.2$\times 10^{-3}$} & \scriptsize{1.50$\times 10^{68}$} &  \scriptsize{2.8$\times 10^{8}$}\\
 \hline
 \end{tabular}
\centering
%\begin{minipage}{0.98\textwidth}
\caption{The gas component parameters for the ring and the ``in'' region of 
the Cartwheel, computed from the summed EPIC-pn datasets, plus relative
densities, volumes, and masses of gas.
\label{gas-flux1}
}
%\end{minipage}
\end{table*}

In spite of the relatively high number of net counts detected from the ring 
(760$\pm$30), the interpretation of the results from spectral
fitting is not straightforward. 
A simple power law fits the data, with a slope of $\Gamma = 2.22 [2.01
- 2.61]$, and a $\chi^2_{\nu} = 0.72(24 dof)$.  While statistically
acceptable, the fit has significant residuals, as already 
remarked in the analysis of
Chandra data (Wolter \& Trinchieri 2004).  
We therefore used the same complex spectral fit
used in Chandra, namely a  \emph{powerlaw} plus \emph{mekal} model, to
account for the excess below 1 keV. Since n$_H$ and kT are degenerate
in the fit, we fixed the parameters to their Chandra values of
n$_H=2.3 \times 10^{21}$ cm$^{-2}$,  $\Gamma = 2.3$,
and  $kT = 0.2$ keV, which are not far from those from a
formal best-fit solution with fixed slope. Only the normalizations of the 
two models were allowed to be free parameters.
The fit has a $\chi^2_{\nu} = 0.75$ (25 dof).
The resulting ``gas'' components has L$_X^{0.5-2}  = 3 \times 10^{40}$ erg/s, as in Chandra, 
while the ``unresolved binary component'' has L$_X^{0.5-2}  = 4.3 \times 10^{40}$ erg/s,
and L$_X^{2-10} = 3.0 \times 10^{40}$ erg/s, which is about a
factor of 2 higher than in the Chandra spectrum. The higher power-law 
contribution relative to Chandra 
might partly be caused by the wings of the
bright sources (especially N.10), which cannot be excluded as
efficiently as in Chandra. The excess is indeed only 10-15\% of the total L$_X^{2-10}$
of the sources detected by XMM-Newton. 
The gas luminosity should be regarded as merely indicative of the true value,
since the Chandra and XMM-Newton extraction regions used in its evaluation
differ.
For the Chandra data, we selected mainly the southwest portion of the ring
and excluded a smaller fraction of the region considered (excluded regions have $\sim 1''$ radius). 
In XMM-Newton, the extraction region encompasses the whole
ring, but the regions used to exclude sources are significantly larger and
cover a large fraction of the brightest (southwest) region of the ring 
itself.
With this caveat in mind, we evaluated the total gas luminosity in
the ring by correcting the XMM-Newton result for the area lost due
to sources, assuming the average surface brightness of the ring, which
probably underestimates the true value, and derived
L$_X^{0.5-2}  = 3.5 \times 10^{40}$ erg/s.

The spectral analysis of the ``in'' region (see Fig.~\ref{gas}) indicated that  two components are needed: 
a plasma model, with a temperature of kT = 0.16 [0.12 - 0.21] keV, similar to the ring 
value, and a power law to account for high energy residuals, which we parameterized with a slope fixed at  $\Gamma = 2$
( $\chi^2_{\nu} = 0.99$ (26 dof)).
A single {\em mekal} fit does not provide a good fit: it has a
$\chi^2_{\nu} > 5$; a single
power law model also does not provide a good fit: it has 
$\chi^2_{\nu} = 1.27$ (27 dof) and shows a bad
distribution of the residuals.
Formally, the luminosities in the two components are  L$_X(0.5-2) 
= 2.9 \times 10^{40} $ erg/s for the gas
and  L$_X(0.5-2/2-10)$  = 2.7/2.8 $\times 10^{40}$ erg/s for the power law.
This latter result is surprising since we do not expect
such a significant contribution from binary sources: HMXBs ignited
by the shock wave should no longer be active and there is no evidence
of a young stellar population as there is in the ring, while LMXBs should 
not yet have formed after the burst.
However, this component could  
instead correspond to a residual background: for instance, we expect a 
contribution from the PSF wings  
of the bright sources in the ring, which have been excluded only partially
(the ring width only includes about 70\% of the encircled energy for
sources in the middle of the ring).
We did not expect to find hot gas in the inner part of the Cartwheel.
It would be tempting to associate the gas to the inner ring or a nucleus,
since the inner ring was found to be bright at IR wavelengths 
(Charmandaris et al. 1999).
However, our data are of insufficient spatial resolution to ascertain
whether the hot gas emission is associated with a particular region of 
the galaxy.

In summary, the total luminosity of the Cartwheel gas
computed using a mean temperature of kT = 0.2 keV,
is $L_X \sim 6.4
\times 10^{40}$ erg s$^{-1}$, about 60\% of which originates in the 
ring alone. 

The mass of this gas is given in Table~\ref{gas-flux1}, separately for the ring and inner regions.  We used the average gas density derived from the spectral parameters assuming that a fully ionized gas (N$_e$ = N$_H$) is distributed in a
volume approximated by a toroid, for
the ring, with 
r$_{in}$ = 25$^{\prime\prime}$, r$_{out}$ = 40$^{\prime\prime}$,
and r$_{torus}$ = 7.5$^{\prime\prime}$, 
plus a cylinder, for the inner gas, with r$_b$ = 25$^{\prime\prime}$, and
h = 15$^{\prime\prime}$. % ($\sim 10^{13}$ pc$^3$).
Overall, the total mass in gas is 
M=$7.5 \times 10^8 \eta^{1/2}$ M$_\odot$ in the Cartwheel.

In a study of the ionized hot gas of a few young galaxies
characterized by an intense star-formation activity, T\"ullmann et al. 
(2006) calculated gas 
masses of about $10^8$ M$_\odot$.
Similar values have also been found, for instance, for the spiral galaxy M83 (M=$2.2 \times 10^8$ M$_\odot$ for
a filling factor of $\eta =1$; Ehle et al. 1998) or the interacting
galaxy NGC 3395 in Arp 270 (M=$1.02 \times 10^8$ M$_\odot$ 
for $\eta=1$; Brassington et al. 2005). In this respect, the Cartwheel 
galaxy appears to have a higher gas mass than
these other young/interacting galaxies. %== QUesta e' corretta pei buchi?  si'.

We estimate also that the thermal energy associated with the hot plasma 
for $\eta=1$
is  $E_X= 3 n_e \eta V kT = 4.6/3.2 \times 10^{56}$ erg for the ring/inner
region.
This again appears to be high, since it is only a factor of 2 lower than the thermal 
energy $E_X = 1.5 \times 10^{57}$ erg  estimated
for the starburst galaxy  NGC3256 (Lira et al. 2002; a spectacular merger, 
one of the most IR luminous system in the local universe).
While caution should be taken when interpreting our results, which depend
on several assumptions about the spectral parameters
and corrections applied (see previous discussion), the high values derived 
could be due to the significant clumpiness of the medium.

Finally, we extract the spectrum of gas in the group within a circular
region of 60$^{\prime\prime}$ radius (see Fig.~\ref{gas}), excluding
the contributions of the galaxies and the background sources.
Given the presence of bright sources in the area, we must again take
into account their contribution outside the excluded regions. In
particular, for the bright source XMM11
to the NE, we used 
an exclusion region of radius $8^{\prime\prime}$. 
A larger radius would reduce too much the area available to
measure the group emission.
The contamination from XMM11 is therefore on the
order of 25\% of its flux. 
We extracted its  spectrum and derived a \emph{powerlaw} index 
$\Gamma = 2.0$, which we added to the model spectrum for
the group emission, described by a \emph{mekal} model, to account for the 
high energy part of the spectrum.  We fixed the normalization of the
power law to be 25\% of the flux of XMM11. 

The resulting group gas has  a temperature kT = $0.21[0.15-0.27]$ keV and a
luminosity  $L_X^{0.5-2} =  9 \times 10^{39}$
erg s$^{-1}$. These values are much lower than typical values of
compact groups (kT $\sim$1 keV and $10^{42}$ erg s$^{-1}$; Ponman et al. 1996,  Mulchaey et al. 2000).
However, observations of dynamically young, spiral-dominated systems
indicate significantly softer and lower $L_X$ IGM (SCG0018-4854: Trinchieri et al. 2008;
HCG16: Belsole et al. 2003; HCG 80: Ota et al. 2004).

The measurements of the gas in the Cartwheel group is formally 
consistent with the extrapolation to low temperatures of the $L_X$ -- $kT$ 
relation for groups and clusters (Mulchaey et al. 2000).
Nevertheless, given the large uncertainties and scatter in the extrapolation,
this result is not conclusive.

\section{IR to UV SED}

%TABELLA SED

\begin{table*}[t]
\centering
\begin{tabular}{|c|cc|cc|cc|cc|cc|cc|}
  \hline
   &  \multicolumn{2}{|c|}{\scriptsize{\emph{V}}} & \multicolumn{2}{|c|}{\scriptsize{\emph{B}}} &
   \multicolumn{2}{|c|}{\scriptsize{\emph{U}}} & \multicolumn{2}{|c|}{\scriptsize{\emph{UVW1}}} &
   \multicolumn{2}{|c|}{\scriptsize{\emph{UVM2}}} & \multicolumn{2}{|c|}{\scriptsize{\emph{UVW2}}} \\
  & \scriptsize{Flux}  & \scriptsize{Mag} & \scriptsize{Flux}  & \scriptsize{Mag}& \scriptsize{Flux}  & \scriptsize{Mag}& \scriptsize{Flux }  & \scriptsize{Mag}& \scriptsize{Flux }  & \scriptsize{Mag}& \scriptsize{Flux }  &
 \scriptsize{Mag} \\
 & \scriptsize{$\times 10^{-12}$}  &  & \scriptsize{$\times 10^{-12}$}  & &
 \scriptsize{$\times 10^{-12}$}  & & \scriptsize{$\times 10^{-12}$}  & &
 \scriptsize{$\times 10^{-12}$}  & & \scriptsize{$\times 10^{-12}$}  &
 \\
\hline \hline
 \scriptsize{\emph{\textbf{N7+N9}}} & \scriptsize{0.93} & \scriptsize{18.37} &  \scriptsize{0.69} & \scriptsize{19.07}  & \scriptsize{0.77} & \scriptsize{18.09}  & \scriptsize{1.26} &  \scriptsize{17.31} &  \scriptsize{1.57} & \scriptsize{17.03} &  \scriptsize{1.82} & \scriptsize{16.94} \\
 \scriptsize{\emph{\textbf{N10}}}  & \scriptsize{0.69} &  \scriptsize{18.68} & \scriptsize{0.57} & \scriptsize{19.27} & \scriptsize{0.63} & \scriptsize{18.32}  & \scriptsize{1.23} & \scriptsize{17.34} &   \scriptsize{1.47} &  \scriptsize{17.12}  & \scriptsize{1.61}  & \scriptsize{17.05}  \\
 \scriptsize{\emph{\textbf{N13+N14}}} & \scriptsize{0.96} & \scriptsize{18.34} &  \scriptsize{0.78} &  \scriptsize{18.94}  &  \scriptsize{0.92}  & \scriptsize{17.91} & \scriptsize{1.66} &   \scriptsize{17.01} &  \scriptsize{2.03} & \scriptsize{16.77} & \scriptsize{2.30} & \scriptsize{16.67}  \\
 \scriptsize{\emph{\textbf{N16+N17}}} & \scriptsize{0.55} & \scriptsize{18.93} &  \scriptsize{0.48} & \scriptsize{19.47}  & \scriptsize{0.49}  & \scriptsize{18.59}  &  \scriptsize{1.09} & \scriptsize{17.45} &   \scriptsize{1.37} &  \scriptsize{17.17} & \scriptsize{1.53} &  \scriptsize{17.08} \\
 \scriptsize{\emph{\textbf{HII region}}} & \scriptsize{2.47} & \scriptsize{17.31} &  \scriptsize{1.41} & \scriptsize{18.03}  & \scriptsize{1.48}  & \scriptsize{17.39} &  \scriptsize{1.91} & \scriptsize{16.67} &  \scriptsize{2.63} & \scriptsize{16.48} & \scriptsize{2.75} & \scriptsize{16.48} \\
 \scriptsize{\emph{\textbf{G1}}} & \scriptsize{6.02} & \scriptsize{16.32} &  \scriptsize{6.08} & \scriptsize{16.72}  & \scriptsize{6.43}  & \scriptsize{15.80} &  \scriptsize{6.55} & \scriptsize{15.52} &  \scriptsize{7.71} & \scriptsize{15.32} &
 \scriptsize{8.20} & \scriptsize{15.29}  \\
  \scriptsize{\emph{\textbf{G2}}} & \scriptsize{10.3} & \scriptsize{15.76} &  \scriptsize{4.73} & \scriptsize{16.99}  & \scriptsize{2.14}  & \scriptsize{17.00} &  \scriptsize{0.66} & \scriptsize{18.01} &  \scriptsize{0.13} & \scriptsize{19.78} &
 \scriptsize{$<$ 0.08} & \scriptsize{$>$20.00}  \\ 
\scriptsize{\emph{\textbf{Ring}}} & \scriptsize{39.7} & \scriptsize{14.29} &  \scriptsize{34.9} & \scriptsize{14.82}  & \scriptsize{29.8}  & \scriptsize{14.13} &  \scriptsize{30.1} & \scriptsize{13.86} &  \scriptsize{36.0} & \scriptsize{13.65} &
 \scriptsize{39.0} & \scriptsize{13.60}  \\
 \hline
\end{tabular}
\centering
%\begin{minipage}{0.93\textwidth}
\caption{Optical and UV flux of ULXs sources and HII region of
the southern ring of the Cartwheel, G1, and the total ring.
 \label{OMflux1} }
%\end{minipage}
\end{table*}

Using data from the OM instrument in the six UV/optical
bands observed, we studied the broad-band energy distribution 
of the ring, of galaxies 
G1 and G2, and of two regions in the ring, corresponding
to the brightest ULX, N.10, and to the brightest IR hot-spot
(labelled HII).  The fluxes obtained in each filter are listed
in Table~\ref{OMflux1}.
We added the IR fluxes from the literature, in particular
the 2MASS data (total magnitude; Skrutskie et al. 2006) for the Cartwheel and G2.
We emphasize that the 2MASS measurement includes the entire Cartwheel,
so they are not strictly comparable to the values reported in Table~\ref{OMflux1}, which refer to the ring only. 
We also included the ring, G1, and G2 values in B and K band from
Marcum et al (1992).

\begin{figure}
\psfig{file=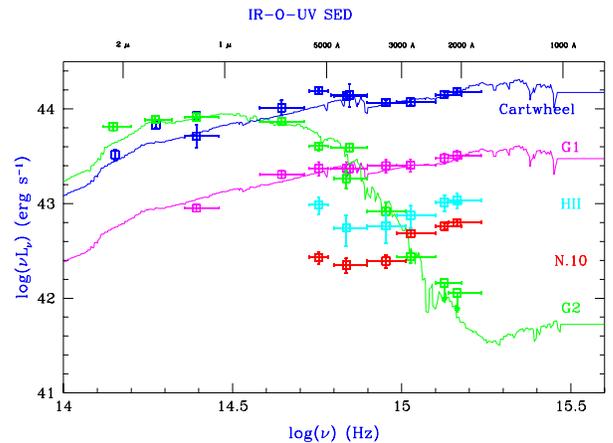,width=9cm,clip=}
%\begin{minipage}{0.9\textwidth}
\caption{SED of Cartwheel ring, G1, G2, and HII and N.10, see
text for description. Data from OM, Marcum et al. 1992, 2MASS
\label{fig_om}}.
%\end{minipage}
\end{figure}

Fig.~\ref{fig_om} shows the Spectral Energy Distributions for the different  sources considered, from Table~\ref{OMflux1} data.  We also show
two models: 1) a model spectrum typical of an old elliptical (generated with the GRASIL code; Silva et 
al. 1998\footnote{see also {\tt 
http://cass.ucsd.edu/SWIRE/mcp/templates/\~swire\_templates.html}}), normalized to the data of G2; 
this model describes the shape of this galaxy spectrum well; 2)
the spectrum resulting from a burst of star formation (courtesy of Paolo Franzetti,
private communication), constructed with the PEGASE code (Rocca-Volmerange \&
Fioc 1999) with a star-formation history from Gavazzi et al. (2002). This model 
describes the Cartwheel ring and, at a slightly lower total
luminosity, G1.
The age of the burst is 400 Myr,
which is comparable to the age of the collision
(300 Myr from Fosbury \& Hawarden 1977, Marcum et al. 1992, and Higdon 1996), while 
the mean stellar age  is about
$\sim 40$ Myr. The ``effective'' optical thickness (related to the extinction
as in Guiderdoni \& Rocca-Volmerange 1987) and the metallicity are computed 
in an auto-consistent way by the model itself.
% tau_lambda = 3.21(1-omeaga_lambda)^1/2 (A_lambda/A_V) (Z_sun) (Zg(t)/Zsun)^s g(t)
The stellar mass of the model is $M = 4.9 \times 10^{9} M_{\odot}$,
which implies that the stars produced in the ring by the shock
represent a small fraction of the total mass of the entire galaxy before the
encounter, estimated to be
$M = 5 \times 10^{11} M_{\odot}$ (Fosbury \& Hawarden 1977).
The normalization of the model provides a measure of the star formation rate
(SFR) corresponding to 
SFR = 23 M$_{\odot}$/yr for the Cartwheel
and SFR = 5 M$_{\odot}$/yr for G1.   
The estimate agrees well with other measures of the
SFR in the Cartwheel (Mayya et al. 2005, Wolter \& Trinchieri 2004).
We note that different combinations of burst age and SFR 
could reproduce the observed spectrum. 
However, the mean stellar age is consistently on the order of a 
few $ 10^7$ yr.
The model also appears to provide an accurate description of G1, 
although it might not be entirely 
appropriate to the evolutionary state of 
G1, since it should not have been involved in the collision.

The HII region and N.10 are similar in shape to the total
Cartwheel ring, if anything even bluer.
However, given the physical size of the extraction regions ($\geq 1$ kpc) 
we expect a large
contamination from the ring in which they are embedded, and it 
is not straightforward to evaluate independently their colors.

\section{Summary}

We have observed the Cartwheel with XMM-Newton at two different
epochs separated by 5 months.
We detected 7 sources in the ring, plus one in the inner part, which
however could be a background source (N.11), 
all at the luminosity level of ULXs. 
The relatively large region ($\sim 6$ kpc) in which these luminosity
measurements were obtained could include a large number of fainter sources,
although from the comparison with Chandra data we know that there are only one
or two dominant individual objects in each detection.
This is also supported by the observed variability of all ULXs:
at least 4 of them have varied in the 4 years between the Chandra
and the first XMM-Newton observation, 
and at least 3 in only 5 months, in addition
to the source N.10 discussed in Wolter et al. (2006).

The variability amplitude and frequency suggest an association
with X-ray binaries.
The optical SED of the Cartwheel ring has very blue colors, 
consistent with a recent burst of SF activity, of age similar to
previous estimates for the tidal encounter that generated the
shock, producing stars of mean age of a few 10$^{7}$ yr.
Binaries are therefore likely to be HMXBs, since the presence of 
a sizable population of IMBHs is unlikely (King 2004).

Overall, the X-ray characteristics appear to be consistent with various
formation and evolutionary models discussed for ring galaxies
and particular the Cartwheel itself (e.g. Mapelli et al. 2008;
King 2004; Marcum et al. 1992; Higdon 1996)
where a radial shock, generated at the impact with the 
companion galaxy, is driven into a predominantly gaseous disk
in which star formation is triggered.
The shock itself, propagating at a speed equal to the expansion velocity of 
the ring ($\sim$ 60 km/s, from HI data; Higdon 1996, or even lower from 
H$\alpha$ measurements; Amram et al. 1998) would heat the gas to a 
temperature of kT $\leq 0.05$ keV, lower than measured.
However, the shock should be responsible for the enhanced
star formation, which in turn heats the gas (Wolter et al. 1999).

The gas of the Cartwheel has a very low temperature and appears to be
quite luminous ($L_X = 6 \times 10^{40}$ erg/s). 
We derive a relatively large gas mass,
both in the ring and in the inner disk region.

We also detect a hot intergroup medium of low temperature.
Although other systems exist with similar low temperatures and luminosities,
the statistical significance of the detection
does not allow us to complete a meaningful comparison with them.

\begin{acknowledgements}

It is always a pleasure to acknowledge useful discussions with
Angela Iovino.  
The analysis of OM data has greatly benefited form the help of
Marcella Longhetti, Mari Polletta, Paolo Franzetti, Ruth
Gr\"utzbach, and Marco Scodeggio.
We thank an anonymous referee for very useful comments that
helped us to improve the paper significantly.
We acknowledge partial financial support from the Italian
Space Agency under contract ASI-INAF I/023/05/0.
This work is based on observations obtained with \emph{XMM-Newton}, 
an ESA science mission with instruments and contributions directly 
funded by ESA Members States and the USA (NASA).
This research has made use of SAOImage DS9, developed by Smithsonian
Astrophysical Observatory.
\end{acknowledgements}

% for the bibliography, at the end
%\bibliographystyle{aa} % style aa.bst
%\bibliography{BIB-ARTICOLO} % your references Yourfile.bib

\end{document}